\documentclass[fleqn,10pt]{wlscirep}
\usepackage[utf8]{inputenc}
\usepackage[T1]{fontenc}

\title{Illuminance-tuned collective motion in fish}

\author[1,+]{Baptiste Lafoux}
\author[1,+]{Jeanne Moscatelli}
\author[1,+,*]{Ramiro Godoy-Diana}
\author[1,+,*]{Benjamin Thiria}
\affil[1]{Laboratoire de Physique et M\'ecanique des Milieux H\'et\'erog\`enes (PMMH), CNRS UMR 7636, ESPCI Paris---PSL Research University, Sorbonne Universit\'e---Universit\'e Paris Cit\'e, 10 rue Vauquelin, 75005 Paris, France}

\affil[*]{ramiro@pmmh.espci.fr, benjamin.thiria@espci.fr}

\affil[+]{these authors contributed equally to this work}

\begin{abstract}
We experimentally investigate the role of illumination on the collective dynamics of a large school (ca. 50 individuals) of \textit{Hemigrammus rhodostomus}. The structure of the group, defined using two order parameters, is quantified while progressively altering the visual range of the fish through controlled cycles of ambient light intensity. We show that, at low light levels, the individuals within the group are unable to form a cohesive group, while at higher illuminance the degree of alignment of the school correlates with the light intensity. When increasing the illuminance, the school structure is successively characterized by a polarized state followed by a highly regular and stable rotational configuration (milling). Our study shows that vision is necessary to achieve cohesive collective motion for free swimming fish schools, while the short-range lateral line sensing is insufficient in this situation. The present experiment therefore provides new insights into the interaction mechanisms that govern the emergence and intensity of collective motion in biological systems. 
\end{abstract}


\begin{document}

\flushbottom
\maketitle
%
%
\thispagestyle{empty}

\section*{Introduction}

Many living systems exhibit fascinating dynamics of collective behavior during locomotion, from bacterial colonies \cite{zhang_collective_2010, gachelin_collective_2014} to human crowds \cite{moussaid_experimental_2009, bain_dynamic_2019} or starlings murmurations \cite{bajec_organized_2009, ballerini_interaction_2008}. These collective motions are characterized by coherent and synchronized displacements on large scales of time and space \cite{chate_modeling_2008}. The emergence of such complex spatio-temporal patterns involving a large number of individuals can be described using local, short-range interactions between nearest neighbors only \cite{katz_inferring_2011}.   

Fish are a typical example of this kind of self-organization: they naturally tend to form ordered groups, called swarms or schools \cite{shaw_schooling_1978}. More than 50\% of fish species school \cite{pitcher_shoaling_1998}, giving an advantage to the group in terms of protection against predators \cite{handegard_dynamics_2012}, foraging \cite{pitcher_functions_1986} or cost of locomotion \cite{weihs_hydromechanics_1973, ashraf_simple_2017}.

From a practical point of view, schooling involves, for each individual in the group, a knowledge of both position in space and kinematics of close neighbors \cite{calovi_disentangling_2018, filella_model_2018}. In order to get this information, fish rely on vision, sensing of hydrodynamic disturbances and chemo-olfactory cues \cite{hemmings_olfaction_1966, pavlov_patterns_2000}. The role of each of these senses is not clearly elucidated today \cite{lopez_behavioural_2012}, but it is generally accepted that vision and hydrodynamic sensing are the most predominant \cite{partridge_sensory_1980, pitcher_fish_1979}.

To sense hydrodynamic disturbances, fish use their lateral line system \cite{bleckmann_lateral_2006}. This ability has been suggested to be a factor in the formation of fish schools \cite{faucher_fish_2010}. It is possible to impair the functioning of the lateral line of fish, resulting in a modified schooling behaviour \cite{faucher_fish_2010, mekdara_effects_2018, mekdara_tail_2021}. However, this kind of invasive procedure may alter the behaviour of the fish in an unexpected manner.

Another way of quantifying the main sensory mechanisms for swimming interaction is to evaluate the role of vision. For instance, the ambient light level can modify the collective response of schooling fish in different situations \cite{puckett_collective_2018, giannini_testing_2020}. Recently, McKee et al.\cite{mckee_sensory_2020} compared the role of the lateral line and vision in schooling fish. They suggested, based on experiments with 5 fish, that although both lateral line and vision are involved in the interaction between individuals, vision should be sufficient for schooling.

Previous studies \cite{partridge_sensory_1980, pitcher_blind_1976} have also addressed the problem of vision with larger schools (20-30 fish), and showed that fish wearing opaque eye covers were able to maintain collective motion, using their lateral line system only. However, in these experiments, only one fish was blinded and placed back in a normal school, which limits the conclusions in terms of collective motion. 

It has been found that fish reduce or completely suppress schooling behavior below a certain light threshold, that can vary across species\cite{whitney_schooling_1969, ryer_effect_1998}. However, these experiments were conducted on 4 to 6 fish and therefore do not provide evidence for specific behaviors that may occur when increasing the number of individuals in the school. Furthermore, the question was tackled in terms of an abrupt limit between a cohesive and a non-cohesive state, without considering the effect of an increase in light level over a wide range once these thresholds are exceeded.

In this work, we go further in addressing the role of vision in the formation of large groups of fish, by altering the vision of all individuals at once. For that purpose, we chose to work with a species of highly cohesive fish, \textit{Hemigrammus rhodostomus}, freely swimming in a large and shallow water tank. The available visual information is gradually altered by modifying the illumination over time, with two cycles of increasing then decreasing ramps. In addition to quantifying the role of vision, our study allows us to evaluate the role of the lateral line in a non-invasive way (typically, the response of fish in an experiment without light informs on their hydrodynamic sensing capabilities). Moreover, the progressive nature of the light variation enables use to fully resolve the transition from non-cohesive to cohesive motion. In contrast with similar previous studies, we worked on schools composed of a large number of individuals (around 50), allowing for a robust statistical analysis of the collective behavior parameters.

The evolution of the parameters characterizing the group cohesion clearly shows that the fish group is unable to organize collectively until the light intensity is sufficient for the fish to see each other. This conclusion is supported by a complete description of the transition from disordered to ordered group dynamics as a function of individual visual capacities.

\section*{Results}

\begin{figure}[!ht]
    \centering
    \includegraphics{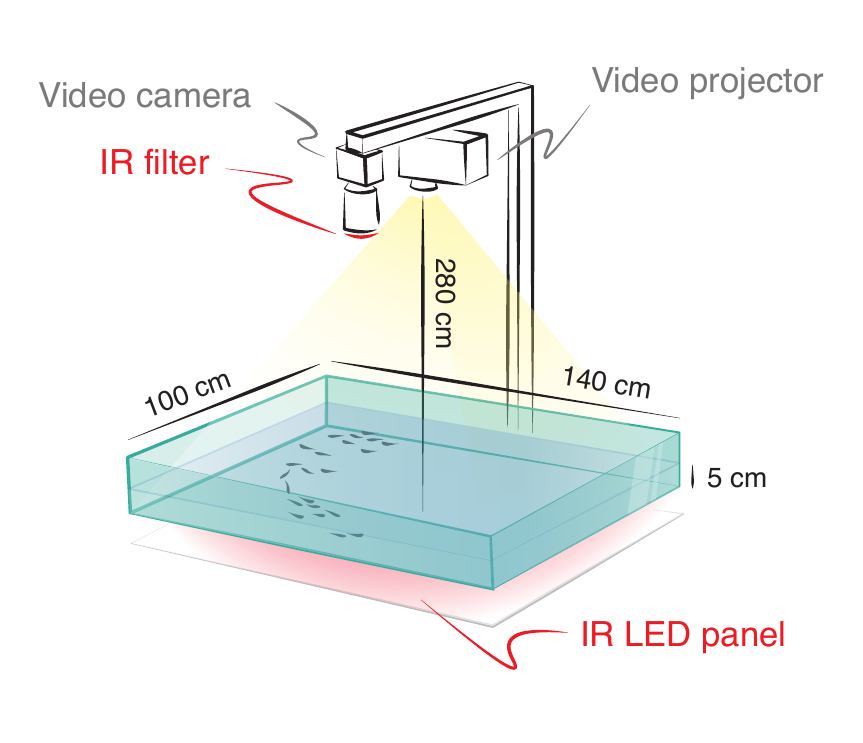}
    \caption{\textbf{Experimental setup} - Groups of fish of around 50 individuals swim freely in a large and shallow tank, while the ambient illumination is continuously modified over time, using a video projector. The whole system is backlit using a custom-build infrared LED panel, and the movements of the fish are recorded with an overhanging camera filming only the IR at 5 frames per second (the visible light is filtered to avoid that its variations alter the lighting conditions of the videos). }
    \label{fig:setup}
\end{figure}

\begin{figure}[!ht]
    \centering
    \includegraphics{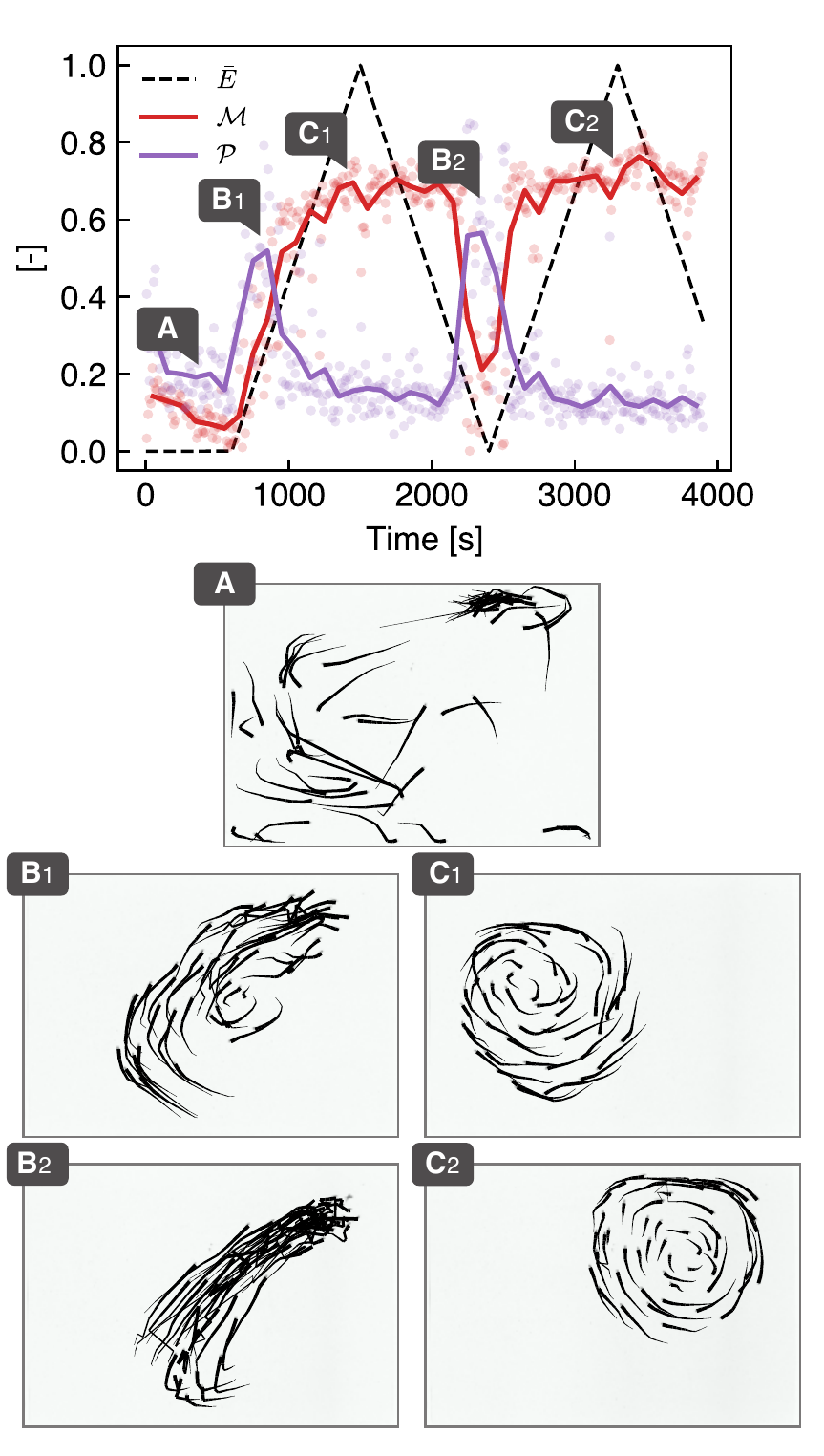}
    \caption{\textbf{The structure of a school of \textit{Hemigrammus rhodostomus} varies when the environment light level is modified over time} - (\textit{top}) Time signal of the order parameters for a group of 53 fish experiencing a variation of normalized illuminance $\bar E$ (Dots : raw signal every 1 s. Lines : signal average with a rolling window of 60 s). $\mathcal P$ is the polarization parameter and $\mathcal M$ the milling parameter. After a 10 min adaptation period in the dark, the group is subjected to slow variations of illuminance, increasing then decreasing, between 0 $\pm$0.1 lx and $E_{\mathrm{max}}$ = 900 lx. (\textit{bottom}) Trajectories snapshots at normalized illuminance $\bar E$ = \{0, 0.22, 0.83, 0.10, 0.89\} (black lines represent trajectories over the last 12 frames, \textit{i.e} 2.4 s).}
    \label{fig:timeserie}
\end{figure}

Our experimental apparatus allows groups of around $N=50$ fish to swim freely in a wide tank while controlling the illuminance $E$ of their environment (see Fig. \ref{fig:setup}). 
During one hour, the ambient illumination is modified over time on a large range (from 0 lux to $E_{\mathrm{max}}$ = 900 lux). The light level is gradually increased and then decreased over a period of 15 minutes in a repeated pattern (2 up-down sequences, see the dashed line in Fig. \ref{fig:timeserie}). 

\begin{figure}[ht!]
    \centering
    \includegraphics{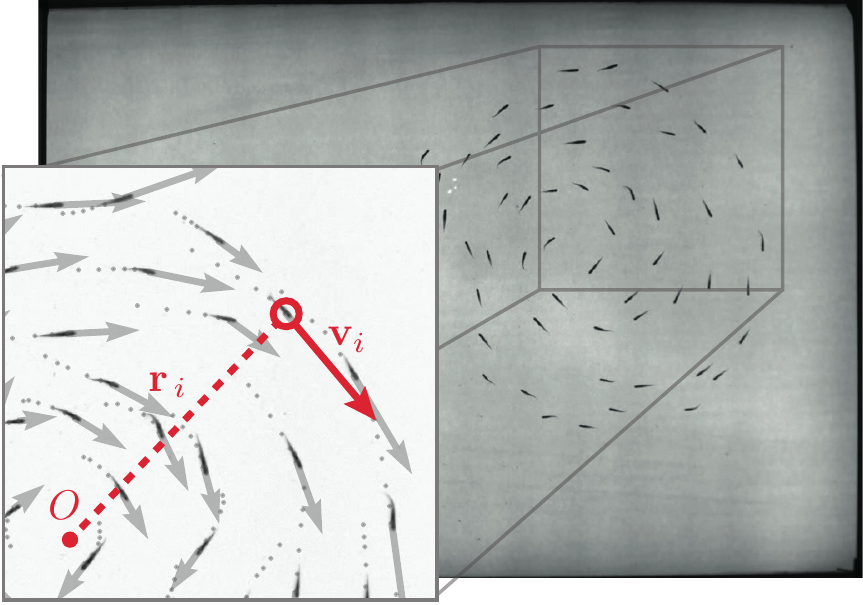}
    \caption{\textbf{Snapshot of 53 fish swimming under illuminance $E$ = 810 lx} - The group motion is captured at 5 fps with a high-resolution camera. From the videos, we reconstruct the individual trajectories of each animal (2D position and velocity at each time step). Zoom-in: 5 previous positions (1s-period, grey dots) and current velocity vectors from tracking data. For the $i$-th fish, we denote $\mathbf v_i$ the instantaneous velocity and $\mathbf r_i$ the position with respect to the school current center of mass $O$.}
    \label{fig:frame_example}
\end{figure}

Fish are recorded using an overhead camera, from which two-dimensional trajectories are extracted with FastTrack, an open-source tracking software \cite{gallois_fasttrack_2021} (see Fig. \ref{fig:frame_example}). In order to quantify the relationship between the fish school organization and the illumination level, we compute two physical quantities that characterize the level of order and cohesion within the group, in terms of alignment and rotation.
The alignment (polarization $\mathcal P$) and rotation (milling $\mathcal M$) order parameters are defined as follows \cite{calovi_disentangling_2018}: 

\begin{equation*}\mathcal P = \left\langle\left|\frac{\mathbf v _i}{\|\mathbf v _i \|}\right|\right\rangle_{i\in 1..N}  \text{, and} \ \mathcal M = \left\langle\left| \frac{\mathbf r_i \times \mathbf v _i}{\|\mathbf r _i \| \|\mathbf v _i \|} \right|\right\rangle_{i\in 1..N, \|\mathbf r _i \| < L}
\end{equation*}
where $\mathbf v_i$ (resp. $\mathbf r_i$) is the instantaneous velocity vector (resp. the position with respect to the school instantaneous center of mass of the $i$-th fish) (see Fig. \ref{fig:frame_example}). $\langle \cdot \rangle$ denotes the averaging operator over all fish in the school. These parameters both range from 0 to 1 and quantify how much the individuals within the school are aligned along the same direction ($\mathcal P$) or rotating around the center of mass of the group ($\mathcal M$). We use a slightly modified expression of $\mathcal M$ to discard cases where the fish are spread over all the tank area and swim along the borders, artificially producing high values of the milling parameter $\mathcal M$, even when the group is not cohesive and no collective milling motion is observed in reality. $\mathcal M$ is thus defined in such a way that we only consider contributions from fish whose distance from the center of mass $\|\mathbf r_i\|$ is less than a threshold value $L$, chosen to be half the short length of the tank ($L$ = 50 cm). 

Additionally, we also quantify two intrinsic characteristic lengths of the fish school: the Nearest-Neighbour Distance (NN-D) and the Inter-Individual Distance (II-D). For a given individual, the NN-D is the distance to the closest fish and the II-D is the average distance to all the other fish. \\

 Variations of illumination strongly influence the values of the milling and polarization paramaters (Friedman test, $\mathcal M$: $\chi^2(8) = 20.07$, $p=1.21\times10^{-3}$, $\mathcal P$: $\chi^2(8) = 22.50$, $p=4.2\times10^{-4}$). Sharp contrasts in behavior are observed, with 3 clearly identifiable phases. On Fig. \ref{fig:timeserie}, we represent a time series of the values of $\mathcal P$, $\mathcal M$, and the normalized illuminance level $\bar E = E/E_{\mathrm{max}}$ (see also Movie S1).

When placed in very dark conditions ($\bar E < 0.05$), fish occupy the entire tank area and move without clear group organization: the average distance between individuals (average II-D) is about 18 body lengths (BL) and fish are placed 1.6 ($\pm$0.2) BL away from their nearest neighbor (average NN-D). Both the rotation parameter and the polarization are very low ($\mathcal M < 0.1$, $\mathcal{P} < 0.2$), showing no significant cohesive movement (Fig. \ref{fig:timeserie}.A).

As the light gradually increases ($\bar E \in [0.05, 0.2]$), a short phase of strong alignment is visible (Fig. \ref{fig:timeserie}.B 1-2), still with a weak but increasing value of the rotation parameter value. Further on, $\mathcal M$ keeps increasing linearly with illuminance, while the polarization drops ($\mathcal P < 0.2$) to eventually reach a plateau for $\bar E > 0.6$ where the behaviour in terms of both rotation and polarization does not change anymore. The school is highly structured, showing a very robust and stable rotational motion ($\mathcal M > 0.6$) with almost no interruption (Fig. \ref{fig:timeserie}.C 1-2). The succession of these phases as a function of illumination is observed repeatedly with great statistical stability, whether the light is following an ascending or descending ramp.

\begin{figure}[ht]
    \centering
    \includegraphics{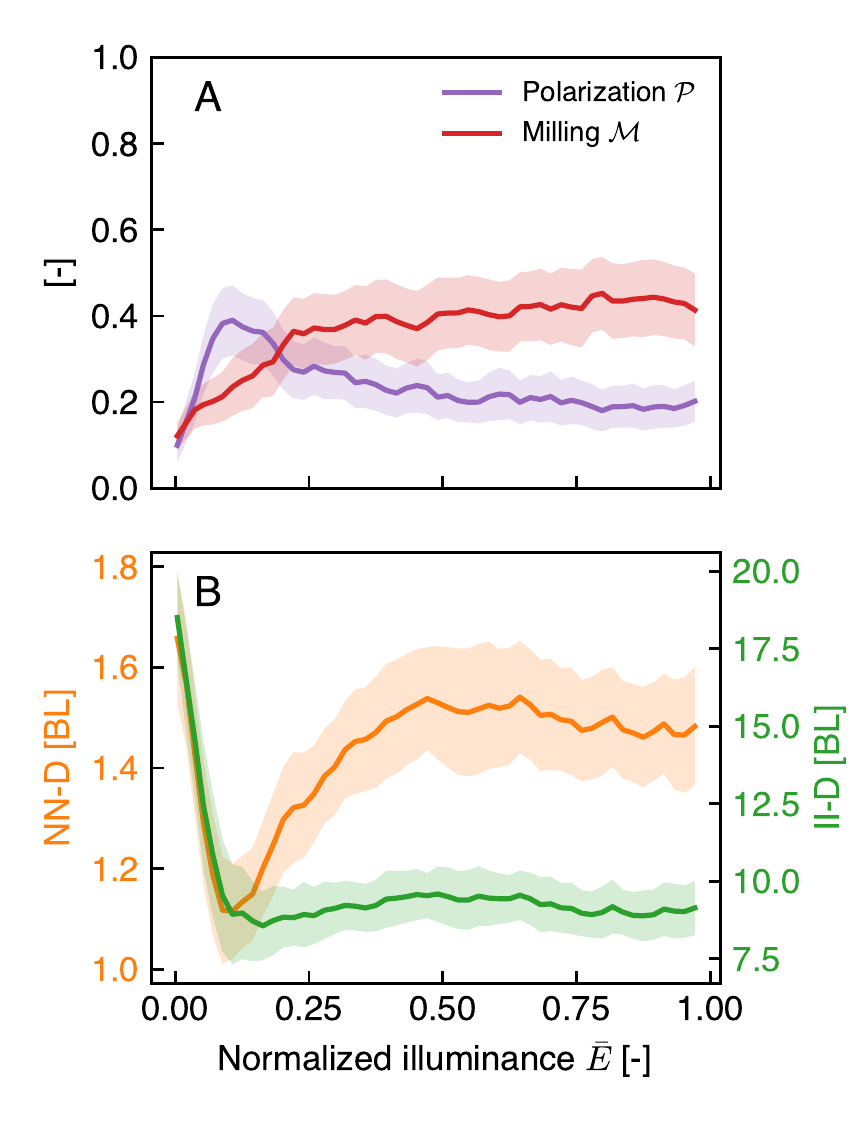}
    \caption{\textbf{Fish school order parameters and distances with respect to light intensity.} Solid lines show values averaged over every trials ($N=6$) and every light ramps (either increasing or decreasing, 4 for each trial), which represents 24 replicates. (The shaded region is the 95\% confidence interval for the mean). \textbf{A.} Polarization and milling parameter. \textbf{B.} Nearest-Neighbour Distance (NN-D) and Inter-Individual Distance (II-D) in Body Lengths. For a given individual, the NN-D is the distance to the closest fish and the II-D is the average distance to all the other fish. Values displayed here are averaged over all individuals in the school. (The scales for NN-D and II-D are different)}
    \label{fig:results}
\end{figure}

Fig. \ref{fig:results} displays the averaged results obtained over 24 different cycles of the illuminance varying experiments (see Materials and Methods). The top graphic (Fig. \ref{fig:results}.A) shows the evolutions of both $\mathcal M$ and $\mathcal P$ with respect to normalized illuminance level $\bar E$. The behaviour described for a single experiment in Fig. \ref{fig:timeserie} is found again in the average curves of Fig. \ref{fig:results}.A: while the milling parameter increases monotonically with illuminance, the polarization parameter peaks rapidly and decreases afterwards to a plateau.  

Fig. \ref{fig:results}.B shows the Nearest-Neighbour Distance (NN-D) and the Inter-Individual Distance (II-D) in Body Lengths (1 BL $\approx$ 3.9 cm). It is worth noting that in its averaged form, the II-D gives a good approximation of the characteristic size of the school.
As can be observed, both characteristic lengths are large when there is no illumination: the average distance between the fish is about 20 BL and the distance to the nearest neighbor is 1.7 BL at the most. This case corresponds to a swarming behavior, without cohesion in the group, as confirmed by Fig. \ref{fig:timeserie}.A. The fish are distributed throughout the tank space and swim independently with respect to each other. 
These quantities rapidly decrease, showing that the individuals within the group get closer to each other as light intensity increases. II-D eventually saturates to a constant value above an illuminance threshold around $E$ = 0.1. As can be observed, the average distance stabilizes around 8 BL, while the NN-D increases progressively before stabilizing around 1.5 BL after $E$ = 0.5, which shows that, within a group of a given size (quantified by the II-D), a better quality of the visual information lets these highly cohesive fish reorganize, finding more regular patterns leaving more space between themselves and their nearest neighbor.

It is known that fish can exhibit changes in their behavior in experiments over time \cite{shearer_experimental_2000}. An additional set of experiments conducted at a fixed illumination level for one hour allowed us to reject the hypothesis that the variations of $\mathcal M$, $\mathcal P$, II-D, and NN-D observed here could be due to the time elapsed since the beginning of the experiment (see Supplementary Material).

\section*{Discussion}

The reading of both Fig. \ref{fig:timeserie} and Fig. \ref{fig:results} is here straightforward. In the absence of light, or with insufficient lighting, fish are unable to give rise to coherent and cohesive group dynamics. We also observed that above a certain threshold, the properties characterizing the collective dynamics do not statistically change with the degree of light intensity and tend to saturate to a constant value. This remark of course holds for the range of illuminance used for this work ($E\in$[0, 900] lx) and the global behaviour of the group might change with higher values of $E$. However, the range used in this work corresponds to lighting values in natural habitats for this kind of animals \cite{fraser_costs_1997}. This sheds light on the recent discussion on the respective roles of vision and lateral line sensing in the appearance of cohesive behaviors. Our observation in the absence of light suggests that lateral line sensing is not sufficient for the group to form a school in free swimming. 

Moreover, the quality of the visual cue seems to be paired with the capacity of the individuals to achieve collective swimming. It is worth noting again that the conclusions brought with this work are based on a large number of individuals constituting the group. This contrasts with most of past studies \cite{morrow_schooling_1948, steven_studies_1959, torisawa_schooling_2007} that characterized cohesion and collective dynamics under a changing illuminance using a reduced group of fish (less than 10 individuals), then mainly focusing on local interactions. Thus, this study constitutes the first experimental work examining vision-based global behavior of a large scale group of fish.

In addition, the stable milling motion observed here with sufficient lighting may in fact be induced by the interaction with walls \cite{tunstrom_collective_2013}. Fig. \ref{fig:results}.A shows that the group polarization starts decreasing after exceeding the visual threshold. This decrease is coupled with the amplification of the milling parameter characterizing the group rotation around its center of mass. Thus, considering that fish tend to align with each other as their ability to see other individuals in the group is enhanced by a brighter environment, the milling behavior could be the simple consequence of being aligned in a confined space. Indeed, the alignment can be either quantified by the polarization or milling parameter, both having the same role in that particular geometry: while the polarization quantifies the alignment along lines, the milling parameters can be understood as a measure of an alignment along circles around the center of the group. The effect of confinement on this transition from polarization to milling is the subject of further investigations.

\section*{Methods}

\subsection*{Fish Breeding}
Rummy-nose tetra (\textit{Hemigrammus rhodostomus}, BL = 3.92 $\pm$ 0.42 cm) were bought from a professional supplier (EFV group, \href{www.efvnet.net}{http://www.efvnet.net}). Fish were kept in a 120L tank on a 14:10 hour photoperiod (day:night), similar to that existing at their latitudes of origin. The water temperature was maintained at 27$^\circ$C ($\pm$ 1$^\circ$C) and fish were fed \textit{ad libitum} with fine pellets from an automated feeder once a day, at a fixed time in the morning. The fish handling protocol complies with the European Directive 2010/63/EU for the protection of animals used for scientific purposes, as certified by the ESPCI Paris Ethics Committee.
\subsection*{Experimental Setup} 
The experimental setup consists of a large shallow tank with a working area of 140 $\times$ 100 cm. This area was illuminated by visible light produced by a video projector (BenQ, 1920 $\times$ 1080 pixels) placed 280 cm above the water surface. To vary the light intensity, uniform images were projected, with shades of gray ranging from 0 to 255 (from complete black to maximum illumination). We measured with a luxmeter that the corresponding light intensity in the tank ranges from 0 to 900 lux ($\pm$3\%). These levels of light intensity are comparable to those existing in the natural environment of origin of the tetra fish, as well as in their breeding conditions. The maximum value $E_{\mathrm{max}}$ = 900 lux corresponds to the illumination on a clear sunny day. 

In order to visualize the fish regardless of the lighting conditions, the tank was lit from the bottom by an infrared LED panel (200W, $\lambda$ = 940 nm, LEDpoint): the wavelength $\lambda$ was chosen to be large enough to be invisible to the fish \cite{carleton_seeing_2020} while not interfering with the light variations created in the environment. The entire device was placed in an enclosure surrounded by opaque curtains, in order to avoid any light or visual disturbance due to the presence of the experimenter. The dimensions of the tank are large enough (25 $\times$ 35 BL) that the interaction of the school with the borders is negligible and the trajectories can be considered unconfined, while the chosen water depth ($h$ = 5 cm) induces swimming movements essentially in 2D without causing stress to the animals.  

The raw data represents 6 hours of video recording at 5 frames per second, with a high spatial resolution (0.7 mm per pixel), which were converted into trajectories (2D position of each fish at each time point) using FastTrack \cite{gallois_fasttrack_2021} tracking software. The tracking accuracy was 94.7\% (percentage of fish detected on average on each frame, compared to the effective number of fish in the tank, see Supplementary Material for details on the tracking accuracy).

\subsection*{Experimental Procedure}
On each experimental run, a group of approximately 50 fish was allowed to swim freely in the tank while the illuminance is varied. The light to which the school was exposed vary slowly enough that the experiment can be considered "quasi-static". In this way, the fish were not stressed by sudden changes in light and the structure of the school changed on a time scale shorter than the light variation.
Video of the full swimming area were recorded with a Basler Camera AC2040-90um monocolor at 5 frames per second, with a resolution of 3 Mpx (2048 x 1536 pixels). A time step represents on average a displacement of 0.15 $\pm$ 0.02 BL. A filter letting only infrared light pass through (IR-transparent PMMA, thickness 3 mm, Lacrylic Shop) was placed in front of the camera lens to enhance the quality of the recorded images. 
The experiments were carried out as follow: first, the light was completely switched off for 5 to 10 min, in order to let the fish get used to the environmental conditions. In this way, an equilibrium regime was reached and the agitation that would be caused by a sudden change of light at the beginning of the experiment was avoided. The illumination then followed two linear growth-decay cycles, from 0 to 900 lux and back. Each ramp lasted 15 min, for an effective experiment duration of 60 min.
We performed 6 such experiments on 3 separate days, for a total of more than 6 hours of video and 24 illuminance ramps studied. Each experiment is performed on approximately 50 fish (56, 55, 53, 49, 58, 50). 
An additional set of 8 experiments at fixed illumination level for one hour was perfomed as a control (see Supplementary Material).
The trials were carried out in sets of two. Animals were randomly selected and then placed in a separate tank after the first experiment. Another group of animals was then chosen for a second experiment, and all were regrouped in the main tank after. This procedure and the large number of bred fish (200) ensures a random distribution of individual across experiments.
The days of experiments were spaced at least 72 hours apart to prevent possible stress, and to avoid conditioning or habituation phenomena. The tests were performed at approximately the same time of day (in the afternoon), to avoid a potential influence of the circadian cycle of the animals on the results. No excess mortality was observed during or in the days following the experiments. 

\subsection*{Statistical Analysis}
A Friedman test \cite{friedman_use_1937} (non-parametric alternative to repeated measurement ANOVA) was conducted to determine whether light intensity had a significant influence on the measured parameters ($\mathcal M$, $\mathcal P$, NN-D, II-D) over the repeated trials. For this test, we binned the data on 8 different intervals of normalized light intensity (regularly spaced, from 0 to 1), with 24 replicates for each of these interval (6 trials, with 4 ramps). We then conducted a post-hoc test to the Freidman test, the Nemenyi test \cite{nemenyi_distribution-free_1963}, to determine the pairwise comparison of means (see Supplementary Material for details). To ensure that the observed behaviors do not depend on the duration of the experiment, a Spearman’s rank correlation was computed to assess the relationship between light intensity and the measures in the continuous experiments, and between elapsed time and the measures in the discrete experiments (See Supplementary Material). There was a high correlation between light and the measured parameters in the first case ($\rho > 0.16$) and a very low correlation between time and the measured parameters in the second case ($\rho < 0.05$).
\bibliography{biblio}

\section*{Acknowledgements}

We thank Amaury Fourgeaud from the Laboratoire de Physique et Mécanique des Milieux Hétérogènes workshop, for his technical support on the experimental setup design.

\section*{Author contributions statement}

B.L. and J.M. carried out the experiments. B.L., R.G.-D. and B.T. designed the experiments, analysed the data and wrote the paper.  All authors participated in discussing and approving the final version of the paper

\section*{Competing interests}
The authors declare no competing interests.

\end{document}